\documentclass[letterpaper, 10 pt, conference]{ieeeconf}  

\IEEEoverridecommandlockouts                              
\usepackage{graphicx}
\usepackage{amssymb}
\usepackage{bm}
\usepackage{amsmath}

\usepackage[colorlinks,allcolors=blue]{hyperref}
\usepackage{mathdots}
\usepackage{svg}
\svgpath{{imgs/}} 
\usepackage{pbalance}

\usepackage{amsthm}
\newtheorem{theorem}{Theorem}
\newtheorem{lemma}{Lemma}
\newtheorem{corollary}{Corollary}
\newtheorem{definition}{Definition}
\newtheorem{assumption}{Assumption}
\newtheorem{conjecture}{Conjecture}
\newtheorem{proposition}{Proposition}
\newtheorem*{remark}{Remark}

\usepackage{fancyhdr}
\title{\LARGE \bf
Adaptive bias for dissensus in nonlinear opinion dynamics with application to evolutionary division of labor games 
}

\author{Tyler M. Paine$^{1,2}$, Anastasia Bizyaeva$^{3}$, and Michael R. Benjamin$^{1}$
\thanks{$^{1}$Department of Mechanical Engineering, Massachusetts Institute of Technology, 
        Cambridge, MA 02139, USA
        {\tt\small tpaine@mit.edu, mikerb@mit.edu}}%
\thanks{$^{2}$Woods Hole Oceanographic Institution
        Woods Hole, MA 02543, USA}%
\thanks{$^{3}$Sibley School of Mechanical and Aerospace Engineering, Cornell University, Ithaca, NY 14853, USA  {\tt\small anastasiab@cornell.edu}}}

\begin{document}

\maketitle
\thispagestyle{empty}
\pagestyle{empty}

\begin{abstract}
This paper addresses the problem of adaptively controlling the bias parameter in nonlinear opinion dynamics (NOD) to allocate agents into groups of arbitrary sizes for the purpose of maximizing collective rewards. 
In previous work, an algorithm based on the coupling of NOD with an multi-objective behavior optimization was successfully deployed as part of a multi-robot system in an autonomous  task allocation field experiment.
Motivated by the field results, in this paper we propose and  analyze a new task allocation model that synthesizes NOD with an evolutionary game framework. 
We prove sufficient conditions under which it is possible to control the opinion state in the group to a desired allocation of agents between two tasks through an adaptive bias using decentralized feedback.
We then verify the theoretical results with a simulation study of a collaborative evolutionary division of labor game.
\end{abstract}

\thispagestyle{fancy}
\section{INTRODUCTION}
In this paper we study the problem of adaptively controlling a bias parameter in nonlinear opinion dynamics (NOD) \cite{Bizyaeva2023NOD} to allocate agents into groups of arbitrary sizes for the purpose of maximizing collective rewards. 
In previous work, an algorithm based on the coupling of opinion dynamics with behavior optimization was successfully deployed as part of a multi-robot system in an autonomous task allocation field experiment \cite{paine2024GCID}. NOD in a dissensus parameter regime was used to allocate agents between two main options in the experiment: exploring for unknown simulated algae blooms or exploiting known blooms by sampling them for confirmation.  The bias in the NOD model -- a key parameter that implicitly defines the allocation ratio for the group -- was heuristically tuned, and performance was measured on the basis of percentage of algae blooms discovered and sampling efficiency.  

\begin{figure}[ht]
  \centering
  \includesvg[inkscapelatex=false, width = \columnwidth]{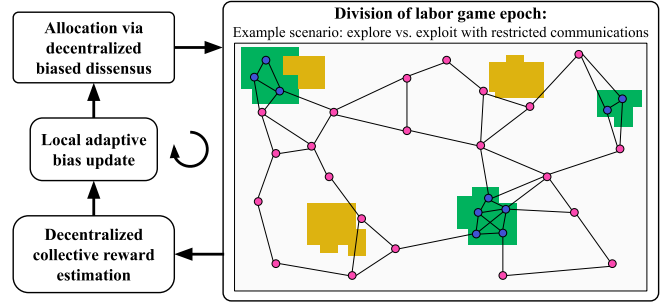}
  \caption{Overview of our approach.  As time progresses the multi-robot system uses biased dissensus to allocate robots in an iterative division of labor game.  An example application to a seek-and-sample scenario is shown where the locations of patches to be sampled change over time.  Robots that chose to explore (pink) cooperatively search for unknown patches (yellow), while sampling robots (blue) cooperatively exploit known patches (green).  Communication is limited by the environment and network connections between robots are shown in black.}
  \label{fig:overview}
  \vspace{-5mm}
\end{figure}

The results from preliminary field experiments of this concept were encouraging, but motivate two theoretical questions we begin to address in this paper.  
First, under what conditions can the bias input be used to precisely control allocation?  Both the network connection graph and bias change as the robots move throughout the environment, complicating the problem of optimal allocation with dissensus. 
Secondly, can an optimal bias be adaptively estimated? 
Many scenarios that could benefit from collective decision-making are complex and the optimal allocation preference is not known by designers a priori.  For this reason it would be advantageous for individual agents to control the bias given some adaptive estimation of the underlying reward structure, all with a theoretical assurance of stability. 

In this paper we show that the dissensus regime of the NOD model is well suited to address the dual objectives of collective estimation and control in decentralized multi-agent autonomous decision making.  Estimating an unknown payoff matrix requires excitation, while controlling the precise allocation of agents with a scalar bias is possible when the network graph has certain structural properties.  We develop these new definitions in this paper, extending recent work on the topic of NOD.  
In several previous applications of NOD the bias is either a static design parameter \cite{Cathcart2023NODHumanMovers} or assumed to be a tunable fixed function, such as the expected payoff in a static public goods game \cite{Park2022TuningGamesNOD}, or the efficiency threshold in resource collection \cite{amorim2023threshold}.  However, an analytical expression for the optimal bias is more difficult to find in complex scenarios where other objectives are considered, such as avoiding collisions and remaining within communication range.  
An alternative approach is to capture the bias preference in a cost functional to be minimized \cite{hu2023emergent}, or to learn a NOD model - including the bias - that is parameterized by a deep neural network \cite{hu2024thinkdeepfastlearning}.  In contrast to these works, the approach described in this paper strikes a balance by maintaining the minimally parameterized structure of the NOD while still enabling adaptation, or learning, of a key parameter that exerts influence on the behavior of the system. 

The NOD framework has been explored in task allocation problems in \cite{Bizyaeva2022Switching} and \cite{amorim2023threshold}. 
In \cite{Bizyaeva2022Switching} the precise allocation ratio of agents across two options, i.e. tasks, is controlled using a signed network structure that is either determined before the deployment or updated locally by individual agents as they deem necessary.
In \cite{amorim2023threshold} agents do not communicate directly, and instead make individual decisions on which task to execute according to a nonlinear opinion update informed by local sensing of their environment. 
Determining the optimal allocation ratio for a group of agents a priori remains a challenge, while correct adjustment during a deployment is also not guaranteed since each agent only has local information about the environment and network, complicating the issue of determining which agent(s) should be assigned to switch. In this study we approach the task allocation problem differently from the previous works by using an adaptive bias to persuade enough of the agents who are the most neutral to join one particular subgroup, eliminating the need to change to the sign of the network connection but taking advantage of inter-agent communication. 
More broadly, although there are many results that prove bifurcation of the equilibrium opinion state occurs \cite{Bizyaeva2023NOD}
\cite{Bizyaeva2021Cascades} \cite{bizyaeva2021Patterns}, preventing deadlocks in the decision process, at the present time there is no analytical framework akin to controllability for the NOD system.  The study in this paper begins to address these gaps.

The primary contributions in this paper are the following. First, we propose a new model of adaptive task allocation that synthesizes the NOD framework and an adaptive evolutionary bias. Second, we present new analysis of the coupled task allocation framework. We prove sufficient conditions on network structure and scalar bias input that enables a group to reach a desired allocation in the NOD framework, and define the persistence of excitation condition in dissensus required for adaptive estimation of the game payoff matrix.  Third, we illustrate effectiveness of the proposed task allocation model in numerical examples.

%

An overview of the approach is provided in Figure \ref{fig:overview} where the example application is a simplified version of the scenario in \cite{paine2024GCID}.  The running example problem in this paper is a large population of agents participating in an evolutionary division of labor game \cite{Du2019DivLaborGames} \cite{cressman2003evolutionary} where the payoff matrix is not known a priori, and the agents make a collective, but biased decision about which option to select. 
The problem formulation is generic and the results can be applied to many scenarios in evolutionary game theory, and other more general scenarios where the reward is quadratic in the population fraction.    

The paper is structured as follows. In Section \ref{sec:def_and_preliminaries} we define notation and some mathematical preliminaries.
As background information, the model of dissensus in NOD \cite{Bizyaeva2023NOD} along with the evolutionary game definition is stated in Section \ref{sec:multi_robot_autonomy_model}. 
In Section \ref{sec:adaptive_bias_formulation} we formulate the control problem present the main theoretical results.
In Section \ref{sec:numerical_simulation} we report numerical simulation studies. 
We conclude in Section \ref{sec:conclusions}.

\section{MATHEMATICAL NOTATION AND PRELIMINARIES}\label{sec:def_and_preliminaries}

Vectors are in bold font, i.e. $\bm{x} \in {\rm I\!R}^n$ with indices $x_i$ for $i = 1,2,\hdots n$.   The $n$ dimensional simplex is $\Delta_n$.  The vector $\bm{1}$ is the vector of all ones. A vector is strictly positive if all elements are positive, and strictly negative if all elements are negative.  We use $\cdot$ to denote multiplication when it adds clarity. The $l_2$ norm is $||.||_2$ and the $l_\infty$ norm is $||.||_\infty$.
The terms agent and robot are used interchangeably, and our target application involves robotic agents.  The set of indices for all robots in the population is $\Omega = {1,2,\hdots,N_a}$ and the cardinality of $\Omega$, $| \Omega | = N_a$. We define $\mathbb{N}$ to be the set of natural numbers.

We describe a network of robots as an undirected graph $G(\Omega, \mathcal{E})$ with set of nodes $\Omega$ and set of edges $\mathcal{E}$.  We assume the graph is strongly connected and $\mathcal{E} = \{ e \ | \ e_{ij} = 1 \ \text{if a connection exists between $\Omega_i$ and $\Omega_j$}   \  i \neq j \}$.   
The degree, or number of nodes connected to node $i$ , is denoted by $\deg(i)$. 
We define $A$ as the adjacency matrix that corresponds to the $G$ without self loops with $a_{ij} = 1$ if $e_{ij} \in \mathcal{E}$.   For the matrix $A$, the smallest eigenvalue is $\lambda^*$ and the associated right unit eigenvector is $\mathbf{v}^*$, normalized to satisfy $||\mathbf{v}^*||_2 = 1$.

The notation $Q(t) \in {\rm I\!R}^{n \times n}$ denotes a doubly stochastic matrix where $[Q(t)]_{ij} = q_{ij}(t) \geq 0$ and $\sum_{j=1}^N q_{ij}(t) = 1 \ \forall i$ and  $\sum_{i=1}^N q_{ij}(t) = 1 \ \forall j$, i.e. the sum of the entries in each row and each column sum to $1$ at every time $t$. The matrix $Q(t)$ is compatible with the undirected graph $G(\Omega, \mathcal{E})$ if $q_{ij}(t) > 0$ for $e_{ij} \in \mathcal{E}$ or $i = j$ and otherwise $q_{ij}(t) = 0$. 
\subsection{Decentralized Dynamic Consensus}

Let $\chi_i \in  {\rm I\!R} $ be the state assigned to agent $i$ in a network of $n$ agents communicating over an undirected graph $G(\Omega,\mathcal{E})$. The network state $\bm{\chi} = \begin{bmatrix} \chi_1 & \chi_2 & \hdots \chi_n \end{bmatrix}^T \in {\rm I\!R}^n$ is updated in discrete-time according to a linear protocol
\begin{equation}
\bm{\chi}(t+1) = Q(t) \bm{\chi}(t). \label{eq:consensus}
\end{equation}
In the following Proposition we state a classic set of sufficient conditions for \eqref{eq:consensus} to define a consensus protocol.
\begin{proposition} [\cite{Garin2010} Theorem 3.2]\label{prop:cons} Consider \eqref{eq:consensus} and assume that $Q(t)$ is doubly stochastic and compatible with a strongly connected graph $G(\Omega, \mathcal{E})$ for all $t \in \mathbb{N}$. Then $\lim_{t \rightarrow \infty} \chi_i(t) = \frac{1}{N}\sum_{i=1}^N \chi_i(0)\  \forall i = 1,2,\hdots n$. \label{prop:consensus}
\end{proposition}
A proof of Proposition \ref{prop:consensus} can be found in \cite{Garin2010} or \cite{seneta2006non}.  In this work we design the matrix $Q(t)$ using the Metropolis-Hastings weights, defined as 
\begin{equation}\label{eq:Q_def}
  q_{ij}(t+1) =  \begin{cases}
    \frac{1}{\max\big( \deg(i), \deg(j) \big) + 1}, & \text{if } (i,j) \in \mathcal{E} \text{ and } i \neq j ,\\
    1 - \sum_{j = 1, i \neq j}^N q_{ij}(t), & \text{if } i = j.
  \end{cases}
\end{equation}
The matrix $Q(t)$ defined through \eqref{eq:Q_def} is doubly stochastic and only requires knowledge of the degree of immediate neighbors \cite{Garin2010}.

\section{BACKGROUND: MULTI-AGENT DECISION AND EVOLUTIONARY GAME MODELS}\label{sec:multi_robot_autonomy_model}

In this section we describe a model of multi-agent decision making using the NOD framework reported in \cite{Bizyaeva2023NOD} along with the definition of the evolutionary division of labor game model used in the running example. These two models are the key ingredients for the adaptive bias setup that will be defined and studied in the remainder of the paper.

\subsection{Nonlinear Opinion Dynamics Model}
Each robot maintains an opinion about two options we will define later in Section \ref{sec:evol_game_model}.  The opinion for the $i^{th}$ robot is captured by the variable $x_i$, and when $x_i \geq  ( < ) \ 0$ the agent will play option $\mathcal{T}_1$ ($\mathcal{T}_2$).
The nonlinear opinion dynamics model developed in \cite{Bizyaeva2021Cascades}, \cite{Bizyaeva2022Switching} and \cite{Bizyaeva2023NOD}, describes the evolution of the opinion as
\begin{equation}
\dot{x}_i = -d x_i + u S \bigg( \alpha x_i + \gamma \sum_{k = 1, k \neq i}^{N} a_{ij} x_{k} \bigg) + b, \label{eq:NOD}
    \end{equation}

where $d > 0$ is the resistance to opinion formation, $\alpha < 0$ and $\gamma < 0$ are the parameters that describe the competitive interaction between own opinions and opinions of other agents, and $S(\cdot)$ is a sigmoid function satisfying $S(0) = 0, S'(0) = 1, S''(0) = 0, S'''(0) \neq 0$.  The attention parameter $u$ is identical on all vehicles and is constant in this study. 
This paper focuses on the scalar adaptive bias $b \in  {\rm I\!R}$ which is described in detail in Section \ref{sec:adaptive_bias_nod}.   

\subsection{Evolutionary Division of Labor Game Model}\label{sec:evol_game_model}
The running example problem in this paper is a mixed strategy evolutionary division of labor game \cite{Du2019DivLaborGames}. 
In this formulation each agent must select one of two options: $\mathcal{T}_1$, and $\mathcal{T}_2$. In the context of an autonomous population, we denote the fraction of the population that plays the $\mathcal{T}_1$ strategy as $y_1$.  The mixed strategy is parameterized  as a vector $\bm{y} = \begin{bmatrix} y_1 &  y_2\end{bmatrix}^T \in \Delta_{1}$ where $y_2 = 1-y_1$. 
The population continuously plays rounds of an evolutionary game with a symmetric payoff matrix 
\begin{align}
\Pi &= \begin{bmatrix}
\pi_{11} & \pi_{12}  \\
\pi_{12} & \pi_{22} \\
\end{bmatrix}, \label{eq:payoff_matrix}
\end{align}

where the traditional replicator dynamics are \cite{cressman2003evolutionary}
\begin{equation}
\dot{y}_i = y_i \big( (\Pi \bm{y})_i - \bm{y}^T \Pi \bm{y} \big).  \label{eq:ref_dynamics}
\end{equation}
In this work we assume the evolutionary game has a strict mixed Nash Equilibrium.  The stability and convergence properties of the strategies is summarized in Proposition \ref{prop:stability_evo_games}

\begin{proposition}[\cite{Cressman2014replicator} Theorem 1]\label{prop:stability_evo_games}
Given an evolutionary game with payoff matrix (\ref{eq:payoff_matrix}) and replicator dynamics (\ref{eq:ref_dynamics}), the following properties are true about the mixed strategy $\bm{y}(t)$:
\begin{enumerate}
\item[i.]  a stable rest point is a Nash Equilibrium. 
\item[ii.] a convergent trajectory in the interior of the strategy space evolves to a Nash Equilibrium 
\item[iii.] a strict Nash Equilibrium is locally stable. 
\end{enumerate}
\end{proposition}

The average observed reward for one iteration of the game is $\mathcal{R}_{ave} = \bm{y}^T\Pi \bm{y}$.   We collect the entries in the payoff matrix (\ref{eq:payoff_matrix}) in a vector $\bm{\pi} = \begin{bmatrix} \pi_{11} & \pi_{12} & \pi_{22} \end{bmatrix}^T$.  Since the parameters in $\bm{\pi}$ enter linearly, we define the regressor vector $\bm{w}(\bm{y}) = \frac{\partial \mathcal{R}_{ave}}{\partial \bm{\pi}}$ by the following relationship
\begin{align}
\mathcal{R}_{ave} = \frac{\partial \mathcal{R}_{ave}}{\partial \bm{\pi}} \bm{\pi} = \bm{w}(\bm{y})^T \bm{\pi}.
\end{align}

\section{MAIN RESULTS: ADAPTIVE BIAS IN DISSENSUS }\label{sec:adaptive_bias_formulation}
In this section we develop the adaptive bias controller with the goal to influence the collective decision-making such that the population state adaptively adjusts the allocation to maximize the payoff.  First we define a sufficient condition akin to controlability on the graph structure, then define a excitation condition for convergence of the estimate of the payoff matrix to to true values.  Finally, we design our adpative bias using the newly developed results. 

In this study the relationship between the opinion $x_i$ of robot $i$ and its strategy is the following: if $x_i \geq (<) \ 0$ then the robot plays strategy $\mathcal{T}_1$ ($\mathcal{T}_2$).  
We define the index sets $\Omega_{\mathcal{T}_1} = \{ i \ | \ x_i \geq 0 \}$ and  $\Omega_{\mathcal{T}_2} = \Omega \backslash \Omega_{\mathcal{T}_1}$.  The mixed strategy observed in the autonomous population at any time $t$ is 
\begin{equation} \label{eq:y_obs}
\bm{y}_{obs} = \begin{bmatrix}
\frac{|\Omega_{\mathcal{T}_1}|}{|\Omega|} &
\frac{|\Omega_{\mathcal{T}_2}|}{|\Omega|}
\end{bmatrix}^T.
\end{equation}

The NOD model (\ref{eq:NOD}) in vector form is 
\begin{equation}
\dot{\bm{x}} = - d \bm{x} + u S\big( ( \alpha \mathcal{I}_{N_a}  + \gamma A) \bm{x} \big) + b\bm{1}, \label{eq:nod_vector}
\end{equation}

and the Jacobian of the system (\ref{eq:nod_vector}) is \cite{Bizyaeva2023NOD}
\begin{equation}
    J = d \mathcal{I}_{N_a} +  u  (\alpha \mathcal{I}_{N_a}  + \gamma A).
\end{equation}

\subsection{Goal of the Adaptive Controller}
The goal is to design an adaptive bias term $b(\bm{x}, \bm{y})$ such that:
\begin{enumerate}
    \item $\lim_{t \rightarrow \infty} \bm{y}_{obs} - \bm{y} < \frac{1}{N_a}$, or the mixed strategy in the population tends to the strict Nash Equilibrium mixed strategy during an infinite sequence games,
    \item the first goal is achieved without knowing the payoff matrix a priori,
    \item all calculations are completely decentralized and only local information is used. 
\end{enumerate}
The goal is to keep the process of robot allocation flexible with regard to which specific agents are assigned to which option.  We seek to design the adaptive bias such that after a sequence of games enough agents are persuaded to join one subgroup such that the ratio approaches the Nash equilibrium mixed strategy.

\subsection{Equilibria Analysis in Biased Dissensus} \label{sec:adaptive_bias_nod}

We first summarize key results that show properties of the nonlinear opinion dynamics with non-zero input $\bm{b}$. 
\begin{proposition}[Competitive Agents \cite{Bizyaeva2023NOD} Corollary IV.1.2.B]\label{cor:BizCompetition}
Consider the nonlinear opinion dynamics (\ref{eq:nod_vector}) with $\alpha \geq 0$, and the associated adjacency matrix $A$.  Let $\lambda^*$ be the smallest eigenvalue of $A$ with the following assumptions: $\lambda^*$ is real, simple and for all other eigenvalues $\lambda \neq \lambda^*$, $Re[\lambda] \neq \lambda^*$.   If $\gamma < 0$, inputs satisfy $b\bm{1}^T\bm{v}^* = 0$, and $\alpha + \lambda^* \gamma > 0$, the model (\ref{eq:nod_vector}) undergoes a supercritical pitchfork bifurcation for 
\begin{align}
u = u_d =& \frac{d}{\alpha + \gamma \lambda^*}
\end{align}
at which opinion-forming bifurcation branches emerge from $\bm{x}=0$.  The associated bifurcation branches are tangent at $\bm{x} = 0$ to  ${\rm I\!R}e({\bm{v}^*})$.  The pitchfork unfolds in the direction given by $b\bm{1}^T\bm{v}^*$, i.e., if $b\bm{1}^T\bm{v}^* > 0 \ (< 0)$, then the only stable equilibrium $\bm{x}^*$ for $u$ close to $u^*$ satisfies $(\bm{x}^*)^T\bm{v}^* > 0  \ (<0)$.
\end{proposition}
 
We summarize the discussion in \cite{Bizyaeva2021Cascades} and \cite{Bizyaeva2023NOD} to highlight our common theoretical foundation.  When $u$ is ``just larger'' than $u^*$ (a point we will clarify later), then non-zero equilibria emerge in subspace of ${\rm I\!R}^{N_a}$ along $V^* = \operatorname{span}\{\bm{v}^*\}$.   The previous statement is true for $u^* < u < u_{\lambda_2}$, where $u_{\lambda_2} = \frac{d}{\alpha + \gamma \lambda_2}$ where $\lambda_2$ is the second smallest eigenvalue of $A$. Given this background, we expand upon previous work with the goal of finding a $b(\bm{y}(\bm{x}))$ such that $\lim_{t \rightarrow \infty} \bm{y}_{obs} - \bm{y} < \frac{1}{N_a}$.  Before we prove our main result, we first state the running assumptions.
\begin{assumption} \label{assump:v_star_properties}
Let $\bm{v}^*$ be the right eigenvector of $A$ corresponding to the uniquely smallest eigenvalue $\lambda^*$ as previously stated.  The graph $G$ and associated adjacency matrix $A$ are such that the vector $\bm{v}^*$ has the following properties:
\begin{enumerate}
	\item \textbf{Unique indices :}\label{assump:v_star_ordered} The entries in $\bm{v}^*$ are unique.  Without loss of generality (\cite{bizyaeva2021Patterns} Theorem 5), there exists a permutation transformation matrix $P$ such that $\bm{v}_P^* = P\bm{v}^*$ where the entries in $\bm{v}_P^*$ are ordered such that
	\begin{equation} \label{eq:v_star_ordering}
	v_{P_1}^* < v_{P_2}^* < v_{P_3}^* < \hdots < v_{P_{N_a}}^*. 
	\end{equation}
    \item \textbf{Orthogonality to bias vector:}, i.e.
        \begin{align}
		(\bm{v}^*)^T \bm{1} = 0. \label{eq:orthog_cond}
		\end{align}
    \item \textbf{Unique linear combination:} Given a constant $\eta_c \in [0,1]$ and $\bm{e}_{J^{-1} \bm{1}}$ is the unit vector for $J^{-1} \bm{1}$.  When any element of the vector sum 
		\begin{align}
			\bm{x}_{sum} = (1-\eta_c) \bm{v}^* + \eta_c \cdot \pm\bm{e}_{J^{-1} \bm{1}} \label{eq:unique_combo}
		\end{align} 
		is equal to zero, no other elements are equal to zero. More precisely, for any index $i$ where $(1-\eta_c) \bm{v}^*_{i} + \eta_c \cdot \pm \bm{e}_{{J^{-1} \bm{1}}_i} = 0$ then $(1-\eta_c) \bm{v}^*_{j} + \eta_c \cdot \pm \bm{e}_{{J^{-1} \bm{1}}_j} \neq 0$ for all $j \neq i$.  The sign of $\bm{e}_{{J^{-1} \bm{1}}_i}$ is such that $\eta_c \in [0,1]$. 
\end{enumerate}
\end{assumption}
All three statements in the assumption are invariant to the (arbitrary) sign of $\bm{v}^*$. 

\begin{remark}
    The properties in Assumption \ref{assump:v_star_properties} can be relaxed if precise allocation is not desired in the entire range of group size $\frac{|\Omega_{\mathcal{T}_1}|}{|\Omega|} = [0, \frac{1}{N_a}, \hdots 1 ]$.  For example, for large networks where the desired allocation ratio is near $0.5$ and the magnitude of excitation is small then properties $1)$ and $3)$ of the entries in $v_*$ at the extreme ends of the set are irrelevant to the performance. 
\end{remark}

Next, we prove a useful lemma and a corollary that set the stage for our main result.

\begin{lemma}\label{lem:interlacing_vstar}
Given the graph $G$ and associated adjacency matrix $A$ are such that Assumption \ref{assump:v_star_ordered} is true.  The agents are reordered with a permutation transformation matrix $P$ such that at entries in $\bm{v}^*$ can be ordered from smallest to largest.  The value zero is interlaced between two elements on this ordered set, i.e. 
\begin{equation} \label{eq:v_star_interlacing}
	v_{P_1}^* < v_{P_2}^* < v_{P_{l-1}}^* \leq 0 < v_{P_{l}}^* < \hdots < v_{P_{N_a}}^*. 
\end{equation}
for some $l$ where $1 < l \leq N_a$. 
\begin{proof}
From \cite{bizyaeva2021Patterns} Theorem 5 the nonlinear opinion dynamics (\ref{eq:nod_vector}) is invariant to the arbitrary permutation of agent order. 
The interlacing result follows from the Perron-Frobenius Theorem (\cite{Farina2000PostiveLinearSystems} Theorem 11) which states that no other eigenvector besides the one corresponding to the largest eigenvalue can be strictly positive.  $\bm{v}^*$ corresponds to the smallest eigenvalue $\lambda^*$ and therefore cannot be strictly positive. 
\end{proof}
\end{lemma}

This lemma provides us with a mechanism to order the robots and therefore the elements of the $\bm{v}^*$ for clarity of notation, as well as introduce the notation of interlacing zero within the set.  
We provide an intermediate corollary to previous work \cite{bizyaeva2021Patterns} to introduce the framework used in our main result. 

\begin{corollary}\label{cor:unbiased_interlacing}
Consider that $\bm{x}^*$ are the stable equilibria of the nonlinear opinion dynamics (\ref{eq:nod_vector}) with $b = 0$ and $u = u^* + \delta_u$ where $\delta_u \in {\rm I\!R}_+$ is an arbitrary small number that satisfies $u^* < u < u_{\lambda_2}$. 
The graph $G$ and associated adjacency matrix $A$ are such that Assumption \ref{assump:v_star_ordered} is true.  
Also consider for clarity of notation that for every time $t$ the system (\ref{eq:nod_vector}) is transformed by the permutation matrix $P$ described in Lemma \ref{lem:interlacing_vstar} such that the value zero is interlaced in the ordered elements of $\bm{v}^*$ at position $l$. 
Then the elements in $\bm{x}^*$ are ordered such that
\begin{equation}
x^*_1 < x^*_2 < \hdots  < x^*_{l-1} \leq  0 < x^*_{l} < \hdots < x^*_{N_a}
\end{equation}
Or equivalently, zero is interlaced in the same position of $\bm{x}^*$ as in $\bm{v}^*$. 
\begin{proof}
The corollary follows from \cite{bizyaeva2021Patterns} Theorem 3.  
\end{proof}
\end{corollary}

We now investigate the properties of the system (\ref{eq:nod_vector}) linearized about the origin and show it is possible to use the input $b\bm{1}$ to interlace $0$ at arbitrary positions within the equilibria set.  
Then we conjecture the same result is true for the nonlinear system (\ref{eq:nod_vector}). 
This main result is given as the following theorem, which is inspired by the more general result on controllabillity for linear networked systems \cite{YangYu2011Controllability}.  The structure of the argument differs from \cite{YangYu2011Controllability} because $|\frac{d}{dt}b| \ll ||\dot{\bm{x}}||_\infty$, i.e. the communication bandwidth is limited such that the bias signal is extremely bandwidth limited and cannot drive $\bm{x}$ to an arbitrary state.  Instead the effect of $b$ is considered as an almost constant bias. 

\begin{theorem}[Opinion Datum Interlacing of Linearized Bounded System]\label{th:dynam_interlacing}
Consider that $\bm{\hat{x}}^*$ are the stable equilibria of the opinion dynamics (\ref{eq:nod_vector}) linearized about the origin,
\begin{equation}
\dot{\bm{\hat{x}}} = \underbrace{((-d + u \alpha)I_{N_a} + u \gamma A)}_J \bm{\hat{x}} + b\bm{1}, \label{eq:NOD_linearized}
\end{equation}
and projected onto the unit sphere, i.e. $||\bm{\hat{x}}||_2 = 1$, with $\frac{-1}{||J^{-1} \bm{1}||_2} < b < \frac{1}{||J^{-1} \bm{1}||_2}$ and $u^* < u < u_{\lambda_2}$. 
Given the graph $G$ and associated adjacency matrix $A$ are such that all statements in Assumption \ref{assump:v_star_ordered} is true.  
Also consider for clarity of notation that for every time $t$ the system (\ref{eq:nod_vector}) is transformed by the permutation matrix $P$ described in Lemma \ref{lem:interlacing_vstar} such that the value zero is interlaced in the ordered elements of $\bm{x}^*$. 
Given these conditions, then the index of the interlaced zero $l_b $, i.e. 
\begin{equation}
\hat{x}^*_1 < \hat{x}^*_2 < \hdots  < \hat{x}^*_{l_{b-1}} <  0 < \hat{x}^*_{l_b} < \hdots < \hat{x}^*_{N_a},\label{eq:dynam_interlacing_expression}
\end{equation}
where $1 < l_b < N_a$ is monotonically decreasing (increasing) by $1$ for monotonically increasing (decreasing) values of $b \in \begin{bmatrix}\frac{-1}{||J^{-1} \bm{1}||_2} &   \frac{1}{||J^{-1} \bm{1}||_2} \end{bmatrix}$ where $\left|\frac{d}{dt}b\right| \ll ||\dot{\bm{x}}||_\infty$.  
\begin{proof}
We decompose $\bm{\hat{x}} = \bm{\tilde{x}} - b J^{-1} \bm{1}$, and for $|\frac{d}{dt}b| \ll ||\dot{\bm{x}}||_\infty$ the bounded linearized dynamics (\ref{eq:NOD_linearized}) are
\begin{align}
\dot{\bm{\tilde{x}}} = J \bm{\tilde{x}}. \label{eq:nod_linearized_coord_trans}
\end{align}
Consider the following
\begin{enumerate}
	\item The magnitude $||\bm{\tilde{x}}||_2$ is a function of $b$. 
		\begin{align}
	||\bm{\hat{x}}||_2^2 = 1 =& ||\bm{\tilde{x}} - b J^{-1} \bm{1}||^2_2 \\
    =&  ||\bm{\tilde{x}}||^2_2 + b^2 \cdot || J^{-1}\bm{1}||^2_2 \nonumber\\
     &+ 2 \cos(\bar{\theta})||\bm{\tilde{x}}||_2 ||b J^{-1} \bm{1}||_2 
	\end{align} where $\bar{\theta}$ is the angle between $\bm{\tilde{x}}$ and $b J^{-1} \bm{1}$.
	\item The bounds on $u$ implies that $d \cdot \alpha + \gamma \lambda_i < 0 \ \forall i = 2,3,\hdots N_a$, and for the system (\ref{eq:nod_linearized_coord_trans}) the stable subspace of ${\rm I\!R}^{N_a}$ is $\operatorname{span}\{\bm{v}_2, \bm{v}_3, \hdots \bm{v}_{N_a} \}$ and the magnitude of all components of the trajectory of $\bm{\hat{x}}$ in the stable subspace decay exponentially to zero. The $\operatorname{span}\{\bm{v}^*\}$ is the unstable subspace. \cite{rugh1993linear}
	\item The condition (\ref{eq:orthog_cond}) implies $(\bm{v}^*)^T(J^{-1} \bm{1}) = (\bm{v}^*)^T \sum_{i=1}^{N_a} \frac{1}{(\alpha -d + \gamma \lambda_i)} \bm{v}_i  \bm{v}_i^T \bm{1} = \frac{1}{(\alpha -d + \gamma \lambda^*)} \bm{v}_i^T \bm{1} = 0$, and therefore $\cos(\bar{\theta}) = 0$. 
	\item The previous three statements imply the $\lim_{t \rightarrow \infty} \bm{\tilde{x}} = \operatorname{sgn}(\bm{x}(t_0)^T \bm{v}^*) \sqrt{1- b^2 \cdot || J^{-1}\bm{1}||^2_2}\bm{v}^*$.   
	\item The equilibrium 
     \begin{align}
     \bm{\hat{x}}^* =& \lim_{t \rightarrow \infty} \bm{\hat{x}} = \operatorname{sgn}(\bm{x}(t_0)^T \bm{v}^*) \sqrt{1- b^2 \cdot || J^{-1}\bm{1}||^2_2 }\bm{v}^* \nonumber \\
     &+ b J^{-1} \bm{1}.  
     \end{align}
	\item Without loss of generality let $\sin(\theta_b)\bm{e}_{J^{-1} \bm{1}} = b J^{-1} \bm{1}$ for $\theta_b \in [-\pi/2, \pi/2]$ since the dynamics are projected onto the $N_a$ sphere and $|| b J^{-1} \bm{1}||_2 \leq 1$. Then 
	\begin{align}
		\bm{\hat{x}}^* =&  \operatorname{sgn}(\bm{x}(t_0)^T \bm{v}^*) \sqrt{1- \sin^2(\theta_b) }\bm{v}^* \nonumber \\ &+ \sin(\theta_b)\bm{e}_{J^{-1} \bm{1}} \\
		\bm{\hat{x}}^* =&  \operatorname{sgn}(\bm{x}(t_0)^T \bm{v}^*) \cos(\theta_b) \bm{v}^* + \sin(\theta_b)\bm{e}_{J^{-1} \bm{1}}.\label{eq:x_hat_star}
	\end{align}
	\item By the condition (\ref{eq:unique_combo}) each index has a unique value of $\eta_c$ where the value equals zero and this corresponds to a unique value of $\theta_b \in [\frac{-\pi}{2},\frac{\pi}{2}]$ where $\frac{(1 - \eta_c)}{\sqrt{(1-\eta_c)^2 + \eta_c^2}} = \operatorname{sgn}(\bm{x}(t_0)^T \bm{v}^*) \cos(\theta_b)$.  The unique constant $b = \frac{\sin(\theta_b)}{||J^{-1}\bm{1}||_2}$. 
    \item \label{st:extreame_linear} When $b =\frac{1}{||J^{-1} \bm{1}||_2}$, the extreme positive limit of the range, 
    \begin{align}
    \bm{\hat{x}}^* &= \frac{1}{||J^{-1} \bm{1}||_2} J^{-1} \bm{1} \\
    &= \frac{1}{||J^{-1} \bm{1}||_2}  \sum_{i=1}^{N_a} \frac{1}{(u\alpha -d + u\gamma \lambda_i)} \bm{v}_i  \bm{v}_i^T \bm{1},
    \end{align}
    for $u^* < u < u_{\lambda_2}$ $(u\alpha -d + u\gamma \lambda_i) < 0 \ \forall \ i$ except for when $\lambda_i = \lambda^*$.  However from the Statement \ref{eq:orthog_cond} the corresponding eigenvector $\bm{v}^*$ is orthogonal to $\bm{1}$. Since $\bm{1}$ is strictly positive, for $b = \frac{1}{||J^{-1} \bm{1}||_2}$  $\bm{\hat{x}}^*$ is strictly negative and therefore $l_b = 1$.  The same argument follows When $b =\frac{-1}{||J^{-1} \bm{1}||_2}$, the extreme negative limit of the range, and we find $\bm{\hat{x}}^*$ to be strictly positive and therefore, with a slight abuse of notation $l_b = N_a+1$. 
\end{enumerate}
We return to the expression of $\bm{\hat{x}}^*$ as a function of $b$ found in (\ref{eq:x_hat_star}).  At the lower limit, values for $b \in [ \frac{-1}{||J^{-1} \bm{1}||_2}, 0]$ map to values of $\theta_b \in [-\frac{\pi}{2}, 0]$.  By the arguments in Statement \ref{st:extreame_linear} when $\theta_b = -\frac{\pi}{2}$ all elements of $\bm{\hat{x}}^*$ are positive and $l_b = 1$. As $\theta_b$  increases monotonically from the lower limit to $0$, the value of $l$ ($N_a - l$) elements in $\bm{\hat{x}}^*$ must cross zero when $\operatorname{sgn}(\bm{\hat{x}}(t_0)^T \bm{v}^*)$ is positive (negative) by inspection of (\ref{eq:x_hat_star}) by Lemma \ref{lem:interlacing_vstar}. Given that  (\ref{eq:unique_combo}) is true each zero crossing occurs at a unique value of $\theta_b$ (and corresponding $b$), and since the ratio of the two functions that scale the vectors is $tan(\theta_b)$, a monotonically increasing function of $\theta_b$, no reversal crossing occurs. Since the assumption  (\ref{eq:unique_combo}) is symmetric about $0$, as $\theta_b$  increases monotonically from $0$ to the upper limit, the same argument regarding the zero crossings applies.  When $\theta_b = \frac{\pi}{2}$ all elements are negative and $l_b = N_a + 1$. 
\end{proof}
\end{theorem}

\begin{figure}[ht]
  \centering
  \includesvg[inkscapelatex=false, width = 0.8\columnwidth]{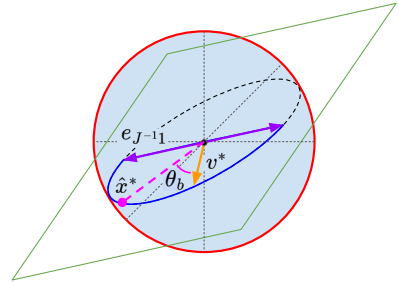}
  \caption{Geometric view of the arguments in Theorem \ref{th:dynam_interlacing}.  The position of the equilibrium point $\bm{x}^*$ lies on a semi-circle arc on a plane that passes through the origin and contains the vectors $\bm{v}^*$ and $J^{-1}\bm{1}$.  The location of $\bm{x}^*$ is parameterized by $\theta_b$ and therefore $b$.  }
  \label{fig:geometric_insight}
\end{figure}

The geometric view of the arguments in this theorem is shown in Figure \ref{fig:geometric_insight}. The condition that zero can be arbitrarily interlaced within elements of a unit vector by a parameter $b$ is equivalent to the vector sweeping a semi-circle  generated by slicing the the $n$ sphere by a 2 dimensional plane passing through the origin. In the geometric case the arc length is parameterized by $\theta_b$ and for the interlacing condition to be true the semi-circle arc must never intersect more than 1 boundary of an orthant at any location. In this case the plane contains the vectors $\bm{v}^*$ and $J^{-1}\bm{1}$, and conditions (\ref{eq:orthog_cond}) and (\ref{eq:unique_combo}) maintain the interlacing condition. 

\begin{conjecture}[Opinion Datum Interlacing of Nonlinear Opinion Dynamics]\label{conj:nod_interlacing}
Given the conditions of Theorem \ref{th:dynam_interlacing} but applied to the nonlinear dynamical system (\ref{eq:nod_vector}), then the index of the interlaced zero $l_b \in \mathcal{N}$, i.e. 
\begin{equation}
x^*_1 < x^*_2 < \hdots  < x^*_{l_{b-1}} <  0 < x^*_{l_b} < \hdots < x^*_{N_a}
\end{equation}
where $1 < l_b < N_a$ is monotonically decreasing (increasing) for monotonically increasing (decreasing) values of $b$.  
\end{conjecture}
This conjecture is supported by the results of Theorem \ref{th:dynam_interlacing}, and should hold true at least for small values of the bias parameter $b$. We leave proving this conjecture and establishing the bounds on its validity to future work. 
For completeness, the issue of bounding all entries in $\bm{x}^*$ above or below zero is addressed in the following lemma. 

\begin{lemma}\label{lem:extreme_inputs}
Consider the stable equilibria $\bm{x}^*$ of the nonlinear opinion dynamics (\ref{eq:nod_vector}) with $u^* < u < u_{\lambda_2}$.  For relatively `large' magnitude positive (negative) input $b$ the resulting stable equilibria  $\bm{x}^*$ has all positive (negative) elements. 
\begin{proof}
At equilibrium $d \bm{x}^* =  u S\big( ( \alpha \mathcal{I}_{N_a}  + \gamma A) \bm{x}^* \big) + b\bm{1} $. We consider the the positive case first.  If $b > u \cdot 1$ then $b \bm{1} > u S\big( ( \alpha \mathcal{I}_{N_a}  + \gamma A) \bm{x}^* \big)$ for every element and therefore $\bm{x}^*$ is strictly positive.  The same argument follows for the negative case. 
\end{proof}
\end{lemma}

\begin{figure*}[ht]
  \centering
  \includegraphics[width = \textwidth]{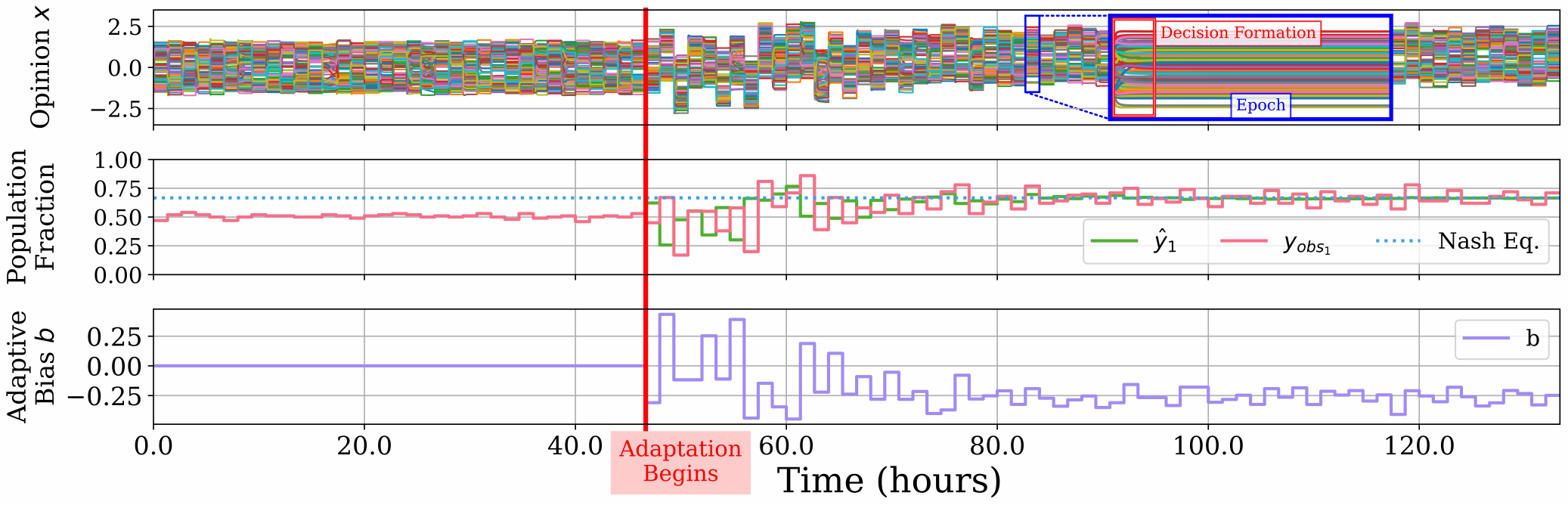}
  \caption{Evolution of the opinion state of 100 networked robots participating in an evolutionary game where each robot plays one of two options depending on the sign of their opinion.  Adaptation begins part way through the simulation, and after a period of excitation required to collectively estimate the underlying payoff matrix the population state approaches the goal state, the Nash Equalibrium, as desired. Watts–Strogatz model with mean degree $K=10$, $\beta=0.1$ \textbf{Top:} Opinion trajectories of all 100 robots where the inset detailed view shows a typical epoch with opinions approaching equilibrium during the initial decision-formation period. $d = 0.5$, $\gamma = -0.03$, $\alpha = 0.3$.  \textbf{Middle:} After adaptation begins the evolution of the observed mixed strategy (\ref{eq:y_obs}) in a system with the adaptive bias (\ref{eq:adaptive_bias_update}) follows the estimate of the mixed strategy Nash Equilibrium via replicator dynamics (\ref{eq:ref_dynamics_update}) which converges to the true Nash Equilibrium.  \textbf{Bottom:} The adaptive bias term during the simulated game iterations.}
  \label{fig:evo_game_adapt_bias}
  \vspace{-4mm}
\end{figure*}

\subsection{Persistent Excitation of Dissensus}\label{sec:PE_dissensus}
As discussed in Section \ref{sec:adaptive_bias_nod} the equilibria state $\bm{x}^*$ changes as the network structure changes.  Although this behavior can present a challenge from the prospective of control, excitation is necessary to discover the underlying reward structure, which in turn improves control.  Thus dissensus is particularly well suited for the task of decentralized autonomous group allocation and exploration. In this section we provide a sequence of definitions to quantify that change, or excitation, that follow from the well known result on the persistent excitation (PE) condition in adaptive systems \cite{narendra_adaptive_book_1989}.

\begin{definition}[PE of Dissensus Bifurcation in NOD]\label{def:PE_v_star}
Given the nonlinear opinion dynamics system (\ref{eq:nod_vector}) with time-varying
network graph $G(\Omega, \mathcal{E}(t))$.  The dissensus bifurcation
is persistently exciting if for piecewise continuous $\bm{v}^*$ there exists $T > 0$ and $\alpha > 0 $ such that 
\begin{equation}
    \int_{t}^{t+T} \bm{v}^*(\tau) [\bm{v}^*(\tau)]^T d \tau  \geq \alpha \mathcal{I}.
\end{equation}
\end{definition}

\begin{definition}[PE of Dissensus Equilibrium State in NOD]\label{def:PE_x_star}
The piecewise continuous agent state equilibrium $\bm{x}^*(t)$ in the nonlinear opinion dynamics system (\ref{eq:nod_vector}) is persistently exciting if there exists $T > 0$ and $\alpha > 0 $ such that 
\begin{equation}
    \int_{t}^{t+T} \bm{x}^*(\tau) [\bm{x}^*(\tau)]^T d \tau  \geq \alpha \mathcal{I}.
\end{equation}
\end{definition}

The sequence of equilibrium states $\bm{x}^*(t)$ for $t\in[t_0, t_0 + T]$ that occur a when a group completes sequence of decisions can be PE by Definition \ref{def:PE_x_star}.  We show an example later in a simulation study.  This excitation can be due to a change in the network structure, as quantified by Definition \ref{def:PE_v_star}, or a change in the bias $b\bm{1}$, or both.   We leave the precise definition to future work.

An essential component of the present study is the definition of the PE condition for our mapping of agent opinion to option selection given the payoff matrix (\ref{eq:payoff_matrix}):
\begin{definition}[PE of State-to-Reward Regressor in NOD]\label{def:PE_pop_state}
The piecewise continuous population state $\bm{w}^T(\bm{y}_{obs}(t))$ in the nonlinear opinion dynamics system (\ref{eq:nod_vector}) is persistently exciting if there exists $T > 0$ and $\alpha > 0 $ such that 
\begin{equation}
    \int_{t}^{t+T} \bm{w}(\bm{y}_{obs}(\tau))  [\bm{w}(\bm{y}_{obs}(\tau)]^T d \tau  \geq \alpha \mathcal{I}.
\end{equation}
\end{definition}
Clearly PE of the equilibrium state (Definition \ref{def:PE_x_star}) is necessary for PE of regressor (Definition \ref{def:PE_pop_state}).

\subsection{Adaptive Estimation of Payoffs}\label{sec:adaptiv_est_payoffs}
This section reports a method for online adaptive estimation of the payoff matrix because in many applications it is not known precisely.  The limitation could be due to an uncertain environment such as density of feeding patches, or unknown effectiveness of a particular strategy when used in large groups, or an interaction of both. 

Again we use a consensus process to generate the estimate of $\mathcal{R}_{ave}$ at the end of each game
\begin{equation}
\bm{\mathcal{R}}(t+1) = Q(t) \bm{\mathcal{R}}(t),
\end{equation}
where $\bm{\mathcal{R}}(0) = \begin{bmatrix} \mathcal{R}_1(0) & \mathcal{R}_2(0) & \hdots \mathcal{R}_n(0) \end{bmatrix}$ are the payoffs realized at the end of each game by each agent.  Using the results from Proposition \ref{prop:cons} we conclude that $\lim_{t \rightarrow \infty} \mathcal{R}_i(t) = \frac{1}{N_a}\sum_{i=1}^{N_a} \mathcal{R}_i(0) = \mathcal{R}_{ave}, \ \ \forall i = 1,2,\hdots N_a $ as a result of the consensus process.  We use this estimate, along with the estimate of the mixed strategy $\bm{y}_{obs}$, as computed in Section \ref{sec:adaptive_bias_def}, to update our estimate of the entries of the payoff matrix. 

The vector $\hat{\bm{\pi}}$ is the estimate of $\bm{\pi}$ with error coordinates $\Delta \pi = \hat{\bm{\pi}} - \bm{\pi}$.  The adaptive update law for $\hat{\bm{\pi}}$ is
\begin{align}
\dot{\hat{\bm{\pi}}} = \eta \bm{w}(\bm{y}_{obs}) \Delta \mathcal{R},
\end{align}
where adaptation gain $\eta \in {\rm I\!R}_+$, $\Delta \mathcal{R} = \bm{w}^T(\bm{y}_{obs}) \hat{\bm{\pi}} - \mathcal{R}_{ave}$, and the true payoff matrix is constant.  It can be shown via Lyapunov's direct method that the error $\Delta \pi$ is globally asymptotically stable with all signals bounded.  Furthermore, if the vector $\bm{w}(\bm{y}_{obs})$ is persistently exciting by Definition \ref{def:PE_pop_state} then it can be shown that $\lim_{t \rightarrow \infty} \hat{\bm{\pi}} = \bm{\pi}$ \cite{narendra_adaptive_book_1989}. 

Given the PE condition in Definition \ref{def:PE_pop_state} is true, as the number of games epochs tends to infinity the replicator dynamics becomes
\begin{align}
\lim_{t \rightarrow \infty} \ \dot{\hat{y}}_i =& lim_{t \rightarrow \infty} \ \hat{y}_i \big( (\hat{\Pi} \bm{\hat{y}})_i - \bm{\hat{y}}^T \hat{\Pi} \bm{\hat{y}} \big) \\ =&  \hat{y}_i \big( (\Pi \bm{\hat{y}})_i - \bm{\hat{y}}^T \Pi \bm{\hat{y}} \big) \label{eq:ref_dynamics_update}.
\end{align}
We can conclude from Theorem \ref{prop:stability_evo_games} statement ii) the convergent trajectory $\bm{\hat{y}}$ evolves to a Nash Equilibrium.  With the assumption that the game has a strict Nash Equilibrium $\bm{y}$, by Proposition \ref{prop:stability_evo_games} statement iii) we conclude that $\bm{\hat{y}}$ is locally stable about $\bm{y}$. 

\subsection{Adaptive Dissensus Bias }\label{sec:adaptive_bias_def}
In this section we use the previous analysis on interlacing (Section \ref{sec:adaptive_bias_nod}), definition of PE (Section \ref{sec:PE_dissensus}) and adaptive estimation (Section \ref{sec:adaptiv_est_payoffs}) to design the adaptive bias. 

The following consensus process provides an estimate for $y_{obs_1}$:
\begin{equation} \label{eq:cons_est_y_obs}
\bm{\xi}(t+1) = Q(t) \bm{\xi}(t),
\end{equation}
where 
\begin{equation}
  \xi_{i}(0) =  \begin{cases}
    1, & \text{if } x_i \in \Omega_{\mathcal{T}_1} \\
    0, & \text{if } x_i \in \Omega_{\mathcal{T}_2}.
  \end{cases}
\end{equation}
We use the result of Proposition \ref{prop:cons} to conclude that $\lim_{t \rightarrow \infty} \xi_i(t) = \frac{1}{N_a}\sum_{i=1}^{N_a} \xi_i(0) = \frac{|\Omega_{\mathcal{T}_1}|}{N_a} = y_{obs_1}, \ \ \forall i = 1,2,\hdots N_a$. 

Using the results of Theorem \ref{th:dynam_interlacing} for the linearized opinion dynamics, and Conjecture \ref{conj:nod_interlacing} for the fully nonlinear system, we design the adaptive population bias term to be 
\begin{align}\label{eq:adaptive_bias_update}
b(t_0)  &= 0 \\
\dot{b} &= \begin{cases}- \eta_0 ( \bar{y}_{obs_1} - \hat{y}_1 )  & \text{if }  \bar{y}_{obs_1} - \hat{y}_1 > \frac{1}{N_a} \label{eq:adapt_bias_law} \\
0 & \text{otherwise,}
\end{cases}
\end{align}
where $\eta_0 \in {\rm I\!R}_+$ is an adaptation gain and $\hat{y}_1$ is defined in Section \ref{sec:adaptiv_est_payoffs}.  

\section{NUMERICAL SIMULATION}\label{sec:numerical_simulation}
In this section  we report two simulated evaluations of the analytical results: 
First we provide a verification of the claim in Conjecture \ref{conj:nod_interlacing}. The second scenario is concerned with long term deployments of large groups that are participating in iterations of an evolutionary game where communication is more limited, and bias updates occur at the end of a game.  In both simulations the sigmoid function $S() = \tanh()$, and the other parameters are provided in the figure captions.

In our simulation studies we considered strongly connected networks generated by three models:  the Erd\H{o}s–R\'{e}nyi model \cite{erdds1959random}, the Watts–Strogatz model \cite{watts1998collective}, and the Barab\'{a}si–Albert model \cite{Albert2002StatisticalNetworks}.  The simulation results were qualitatively similar across all three network graph types.  Only one example is shown due to page constraints. 

\subsection{Monotonically Increasing Bias}
\begin{figure}[ht]
  \centering
\includegraphics[width = \columnwidth]{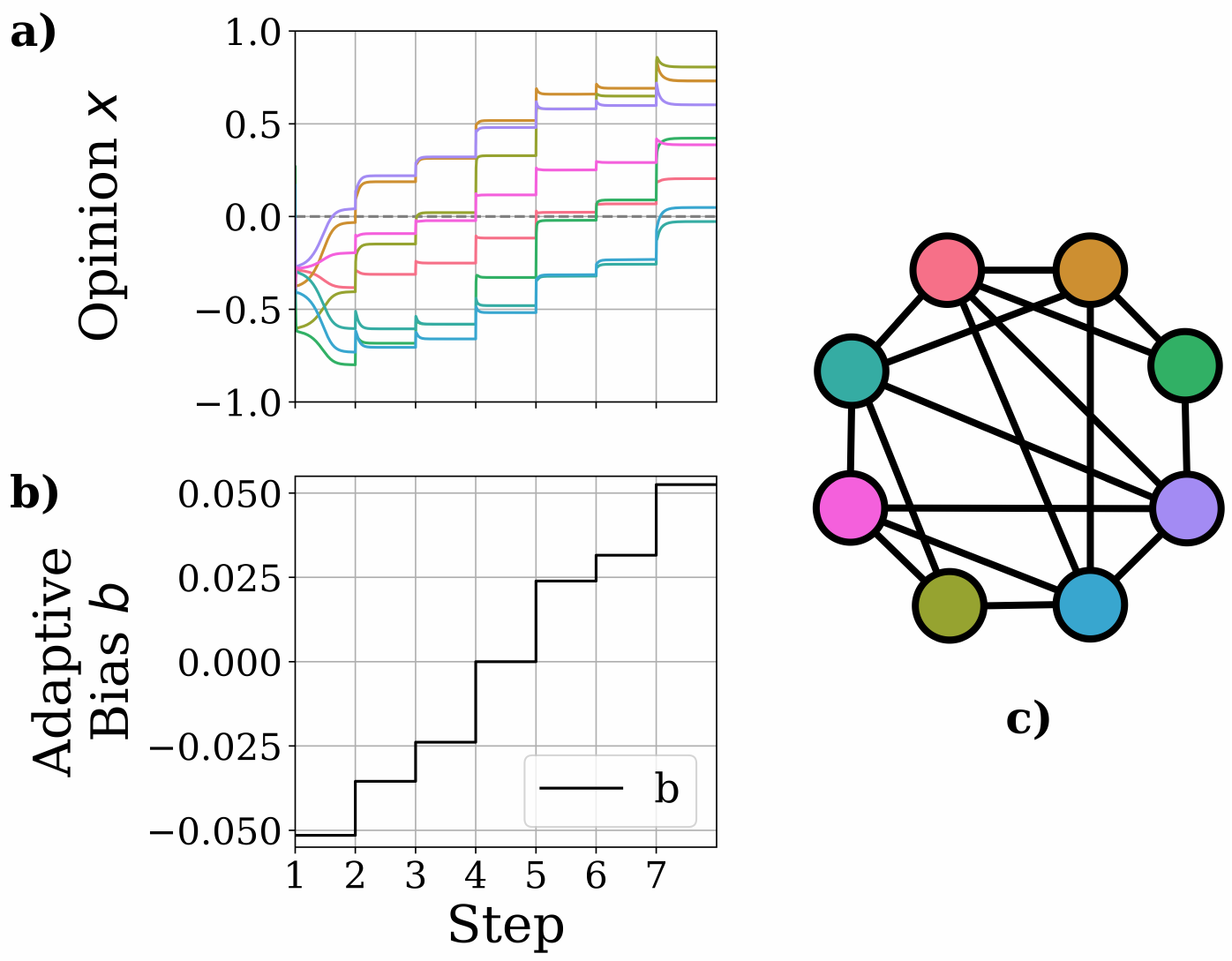}
  \caption{Simulated demonstration of Theorem \ref{th:dynam_interlacing} and Conjecture \ref{conj:nod_interlacing} for the nonlinear system (\ref{eq:nod_vector}). Eight robot nodes in a randomly generated Barab\'{a}si–Albert network ($m=2$) are subject to a monotonically increasing bias $b$.  \textbf{a)} The opinion trajectories cross zero one at a time as bias increases.  \textbf{b)} Bias $b$ step increases. \textbf{c)} Randomly generated network topology,  $d = 0.5$, $\gamma = -0.03$, $\alpha = 0.3$} 
  \label{fig:interlacing_8}
\end{figure}

We show an example of the claim in Conjecture \ref{conj:nod_interlacing}. The bottom plot in Figure \ref{fig:interlacing_8} shows the step increases in bias $b$ and the corresponding opinion state is plotted on the top.  For this evaluation we first calculated the step intervals using the approach described in Theorem \ref{th:dynam_interlacing} and then adjusted some intervals for the fully nonlinear system by $<20\%$ for good visualization.  Overall, the simulated results agreed with the analysis.

\begin{figure}[ht]
  \centering
\includegraphics[width = \columnwidth]{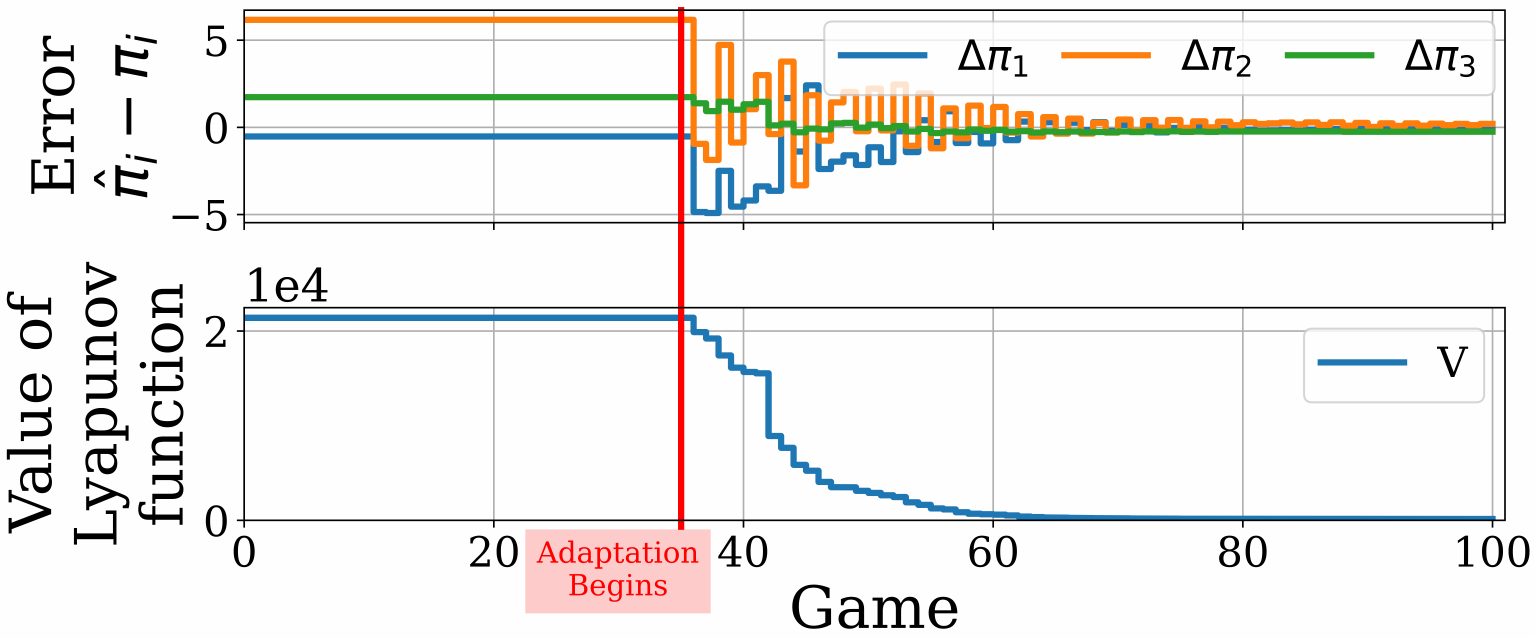}
  \caption{Evolution of errors in estimate of the payoff matrix during iterations of the evolutionary game in Figure \ref{fig:evo_game_adapt_bias}.  Top: Error in estimate of each entry tends to zero after adaptation begins.  Bottom: Value of the Lyapunov function asymptotically tends to zero after adaptation begins.}
  \label{fig:evo_game_errors}
  \vspace{-5mm}
\end{figure}

\subsection{Application to Population Games}
We consider a large group of $N_a = 100$ robots participating in a scenario that can be modeled as an evolutionary division of labor game where the symmetric payoff matrix - and thus the best mixed strategy - is not known precisely a priori.  As the game progresses each robot uses the decentralised approach described herein to update their estimate of the payoff matrix, their goal mixed strategy for the next round, and their dissensus bias.  Finally, since it is common in field deployments for robots to become disconnected with some members of the group, we generate an entirely new graph at the start of each game iteration.  This assumption is rather conservative since some continuity of communication can be expected, but the fact that we find good performance despite the complete reset of connectivity further demonstrates the robustness of the approach.

The top plot of Figure \ref{fig:evo_game_adapt_bias} shows the evolution of each of the 100 robots opinion over the course of 100 games, each with a new graph structure.  Although it is impossible to present all 100 trajectories clearly, the middle plot shows the evolution of the population fraction, or the mixed strategy as the iterations progress.  The value of the adaptive bias is shown in the bottom plot.  Both $y_{obs_1}$ and $b$ do not asymptotically converge to constant values because at every iteration a new random strongly connected graph is generated. 
The estimation errors and associated Lyapunov function are plotted in Figure \ref{fig:evo_game_errors}. 
These results confirm the analytical results: as the game progresses the error in the estimate of the payoff matrix asymptotically approaches to zero and the allocation of the population, or mixed strategy, tends to the true Nash Equilibrium. 

We present this result with the following caveat: The value of $\lambda^*$ cannot be precisely calculated for large matrices due to limitations of numerical precision. Instead we used $u = 1.1 u^*$ in our simulations.  

\section{CONCLUSION}\label{sec:conclusions}
In this paper we proposed and analyzed a task allocation framework based on the NOD model \cite{Bizyaeva2023NOD}. We established conditions for both the collective estimation of rewards and the control of bias in decision-making in large groups in the NOD framework.  Using these results, a new adaptive controller was designed to estimate the payoff matrix in an evolutionary game and adjust the bias of the NOD model with the goal of allocating the group into two subgroups to maximize the collective payoff.  The approach is decentralized, only requires a strongly connected graph structure, and applies to populations of arbitrary size.  Future work includes extending this analysis to scenarios with more than two options and validating this approach in field testing. 
 
\bibliographystyle{IEEEtran}
\bibliography{IEEEabrv,refs_pop_auton}

\begin{thebibliography}{10}
\providecommand{\url}[1]{#1}
\csname url@samestyle\endcsname
\providecommand{\newblock}{\relax}
\providecommand{\bibinfo}[2]{#2}
\providecommand{\BIBentrySTDinterwordspacing}{\spaceskip=0pt\relax}
\providecommand{\BIBentryALTinterwordstretchfactor}{4}
\providecommand{\BIBentryALTinterwordspacing}{\spaceskip=\fontdimen2\font plus
\BIBentryALTinterwordstretchfactor\fontdimen3\font minus \fontdimen4\font\relax}
\providecommand{\BIBforeignlanguage}[2]{{%
\expandafter\ifx\csname l@#1\endcsname\relax
\typeout{** WARNING: IEEEtran.bst: No hyphenation pattern has been}%
\typeout{** loaded for the language `#1'. Using the pattern for}%
\typeout{** the default language instead.}%
\else
\language=\csname l@#1\endcsname
\fi
#2}}
\providecommand{\BIBdecl}{\relax}
\BIBdecl

\bibitem{Bizyaeva2023NOD}
A.~Bizyaeva, A.~Franci, and N.~E. Leonard, ``Nonlinear opinion dynamics with tunable sensitivity,'' \emph{IEEE Transactions on Automatic Control}, vol.~68, no.~3, pp. 1415--1430, 2023.

\bibitem{paine2024GCID}
T.~M. Paine and M.~R. Benjamin, ``A model for multi-agent autonomy that uses opinion dynamics and multi-objective behavior optimization,'' in \emph{2024 IEEE International Conference on Robotics and Automation (ICRA)}, 2024, pp. 8305--8311.

\bibitem{Cathcart2023NODHumanMovers}
C.~Cathcart, M.~Santos, S.~Park, and N.~E. Leonard, ``Proactive opinion-driven robot navigation around human movers,'' in \emph{2023 IEEE/RSJ International Conference on Intelligent Robots and Systems (IROS)}, 2023, pp. 4052--4058.

\bibitem{Park2022TuningGamesNOD}
S.~Park, A.~Bizyaeva, M.~Kawakatsu, A.~Franci, and N.~E. Leonard, ``Tuning cooperative behavior in games with nonlinear opinion dynamics,'' \emph{IEEE Control Systems Letters}, vol.~6, pp. 2030--2035, 2022.

\bibitem{amorim2023threshold}
G.~Amorim, M.~Santos, S.~Park, A.~Franci, and N.~E. Leonard, ``Threshold decision-making dynamics adaptive to physical constraints and changing environment,'' in \emph{2024 European Control Conference (ECC)}, 2024, pp. 1908--1913.

\bibitem{hu2023emergent}
H.~Hu, K.~Nakamura, K.-C. Hsu, N.~E. Leonard, and J.~F. Fisac, ``Emergent coordination through game-induced nonlinear opinion dynamics,'' in \emph{2023 62nd IEEE Conference on Decision and Control (CDC)}, 2023.

\bibitem{hu2024thinkdeepfastlearning}
\BIBentryALTinterwordspacing
H.~Hu, J.~DeCastro, D.~Gopinath, G.~Rosman, N.~E. Leonard, and J.~F. Fisac, ``Think deep and fast: Learning neural nonlinear opinion dynamics from inverse dynamic games for split-second interactions,'' 2024. [Online]. Available: \url{https://arxiv.org/abs/2406.09810}
\BIBentrySTDinterwordspacing

\bibitem{Bizyaeva2022Switching}
A.~Bizyaeva, G.~Amorim, M.~Santos, A.~Franci, and N.~E. Leonard, ``Switching transformations for decentralized control of opinion patterns in signed networks: Application to dynamic task allocation,'' \emph{IEEE Control Systems Letters}, vol.~6, pp. 3463--3468, 2022.

\bibitem{Bizyaeva2021Cascades}
A.~Bizyaeva, T.~Sorochkin, A.~Franci, and N.~Ehrich~Leonard, ``Control of agreement and disagreement cascades with distributed inputs,'' in \emph{2021 60th IEEE Conference on Decision and Control (CDC)}, 2021, pp. 4994--4999.

\bibitem{bizyaeva2021Patterns}
A.~Bizyaeva, A.~Matthews, A.~Franci, and N.~E. Leonard, ``Patterns of nonlinear opinion formation on networks,'' in \emph{2021 American Control Conference (ACC)}, 2021, pp. 2739--2744.

\bibitem{Du2019DivLaborGames}
J.~Du, ``An evolutionary game coordinated control approach to division of labor in multi-agent systems,'' \emph{IEEE Access}, vol.~7, pp. 124\,295--124\,308, 2019.

\bibitem{cressman2003evolutionary}
R.~Cressman, \emph{Evolutionary dynamics and extensive form games}.\hskip 1em plus 0.5em minus 0.4em\relax MIT Press, 2003, vol.~5.

\bibitem{Garin2010}
F.~Garin and L.~Schenato, \emph{A Survey on Distributed Estimation and Control Applications Using Linear Consensus Algorithms}.\hskip 1em plus 0.5em minus 0.4em\relax London: Springer London, 2010, pp. 75--107.

\bibitem{seneta2006non}
E.~Seneta, \emph{Non-negative matrices and Markov chains}.\hskip 1em plus 0.5em minus 0.4em\relax Springer Science \& Business Media, 2006.

\bibitem{Cressman2014replicator}
R.~Cressman and Y.~Tao, ``The replicator equation and other game dynamics,'' \emph{Proceedings of the National Academy of Sciences}, vol. 111, no. supplement\_3, pp. 10\,810--10\,817, 2014.

\bibitem{Farina2000PostiveLinearSystems}
L.~Farina and S.~Rinaldi, \emph{Definitions and Conditions of Positivity}.\hskip 1em plus 0.5em minus 0.4em\relax John Wiley \& Sons, Ltd, 2000, ch.~2, pp. 7--15.

\bibitem{YangYu2011Controllability}
Y.-Y. Liu, J.-J. Slotine, and A.-L. Barabasi, ``Controllability of complex networks,'' \emph{Nature}, vol. 473, pp. 167--73, 05 2011.

\bibitem{rugh1993linear}
W.~Rugh, \emph{Linear System Theory}, ser. Prentice-Hall information and system sciences series.\hskip 1em plus 0.5em minus 0.4em\relax Prentice Hall, 1993.

\bibitem{narendra_adaptive_book_1989}
K.~S. Narendra and A.~M. Annaswamy, \emph{{S}table {A}daptive {S}ystems}.\hskip 1em plus 0.5em minus 0.4em\relax Prentice-Hall, 1989.

\bibitem{erdds1959random}
P.~Erd\H{o}s and A.~R\'{e}nyi, ``On random graphs {I},'' \emph{Publicationes Mathematicae}, vol.~6, no. 290-297, p.~18, 1959.

\bibitem{watts1998collective}
D.~J. Watts and S.~H. Strogatz, ``Collective dynamics of ‘small-world’ networks,'' \emph{Nature}, vol. 393, no. 6684, pp. 440--442, 1998.

\bibitem{Albert2002StatisticalNetworks}
R.~Albert and A.-L. Barab\'asi, ``Statistical mechanics of complex networks,'' \emph{Rev. Mod. Phys.}, vol.~74, pp. 47--97, Jan 2002.

\end{thebibliography}

\end{document}